\documentclass[prl,twocolumn,superscriptaddress]{revtex4}
\usepackage{graphicx}
\usepackage{url}

\begin{document}

\title{Copper Tellurium Oxides - A Playground for Magnetism}

\author{M. R. Norman}
\affiliation{Materials Science Division, Argonne National Laboratory, Argonne, IL 60439}

\begin{abstract}
A variety of copper tellurium oxide minerals are known, and many of them exhibit either unusual
forms of magnetism, or potentially novel spin liquid behavior.  Here, I review a number of
the more interesting materials with a focus on their crystalline symmetry and, if known, the nature of their
magnetism.  Many of these exist (so far) in mineral form only, and most have yet to have their
magnetic properties studied.  This means a largely unexplored space of materials 
awaits our exploration.
\end{abstract}

\date{\today}

\maketitle

In 2005, Dan Nocera's group reported the synthesis of the copper hydroxychloride mineral, herbertsmithite \cite{shores}.
A number of years later, they were able to report the growth of large single crystals \cite{chu}.  Since then, a number
of relatives of this mineral have been discovered and characterized \cite{RMP}.  Despite the existence of a large Curie-Weiss
temperature of order 300 K, herbertsmithite does not order down to 20 mK \cite{mendels}.

The reason these events have
significance is that these minerals could be a realization of an idea proposed by Phil Anderson
back in 1973 \cite{phil} that was based on an early debate in the field of magnetism between Louis N\'{e}el and Lev Landau.
N\'{e}el had proposed the existence of antiferromagnetism, where there are two sub lattices of ferromagnetic moments
oppositely aligned.  This state was subsequently seen by neutron scattering (which resulted in a Nobel prize for  N\'{e}el,
and later for the neutron scatterer, Clifford Shull).  But at the time, there was great skepticism about the existence of this state.
The reason is that it is not an eigenstate of the spin operator (unlike ferromagnetism).  There was suspicion that the
true ground state would be a singlet.  We now know that the origin of the N\'{e}el state is broken symmetry \cite{phil52}, and that
fluctuations are usually not enough to destabilize long range order.
But Phil realized that if the spins sat on a non-bipartite lattice, matters could change.  Imagine a triangle with Ising spins.
Then if two spins are anti-aligned, the direction of the third spin is undetermined.  Phil speculated that instead of N\'{e}el order,
the spins instead paired up to form singlets, and this would be preferred in two dimensions 
(where thermal fluctuations have a tendency to suppress order) and for low spin (where quantum fluctuations 
are more important).  This is particularly obvious for S=1/2, where a singlet bond has an energy of -3$J$/4
compared to -$J$/4 for an antiferromagnetic bond.  But to avoid the energy loss from the unpaired spin, these singlets
should fluctuate from bond to bond, much like Pauling's model for how double carbon bonds in benzene rings resonate from one link to the
next (hence the name, resonating valence bonds).

Most attention has been given to the Heisenberg model, given the more important role of fluctuations in this case.
But we now know that the near neighbor Heisenberg model on a
triangular lattice does order, with the spins rotating by 120$^\circ$ from one sub lattice to the next \cite{huse}.  This is most clear from exact
diagonalization studies, where precursors of the broken symmetry state, and associated magnon excitations, are evident in the eigenvalue
spectrum \cite{bernu}.  But for the kagome case, the spectrum is qualitatively different, with no signature of these effects \cite{lech}.   Over the
years, a number of numerical studies have been done, either purporting a valence bond solid (an ordered array of singlets), or various types
of quantum spin liquid states (gapped Z$_2$, gapless U(1), chiral, etc., where the group corresponds to an emergent gauge group associated
with the symmetry of the spin liquid).  The uncertainty is connected to the fact that all of these states have energies comparable to one
another.  The real interest, though, is that these solutions are characterized by fractionalized excitations (typically free spin 1/2 neutral fermions
known as spinons, or gauge flux excitations known as visons).  Proving the existence of such excitations is a major challenge in physics \cite{balents,savary}.

This brings us to real materials.  Many of them either have contributions over and beyond that of a near neighbor Heisenberg model (longer
range exchange, anisotropic exchange, Dzyaloshinskii-Moriya interactions, etc.) or distorted lattices, all of which can in principle either
change the nature of the ground state, or promote or destabilize order.  Hence the interest in finding as many materials as possible that have frustrated lattices for their magnetic ions, and then characterizing these materials.  In that context, we can often allow Mother Nature to do
the hard work for us.  Many minerals are known which have the appropriate magnetic lattices, making them ideal sources for finding desired materials.
In fact, Dan Nocera's group had first synthesized the iron jarosite minerals (where the iron ions sit on a kagome lattice), but realized that
the same would not apply to a copper (S=1/2) version, since Cu$^{2+}$ would not go onto the Fe$^{3+}$ site.  Hence the turn to
herbertsmithite once they had seen that structure in the mineralogical literature.  In that context, there is one mineral where copper goes
into a jarosite-type structure, osarizawaite \cite{osariza}, but in this case, the copper kagome lattice is strongly diluted by Al$^{3+}$ ions.
The magnetic properties of this mineral, along with a host of others, are unknown at this time.

Last year, in searching the mineralogical literature, I became aware of a review article on tellurium oxides \cite{christy}.  Copper (desired
since it is an S=1/2 ion) has a tendency to be associated with tellurium, and indeed there are many interesting copper tellurium oxides
(and related hydroxides and hydroxyhalides) that can be found in this article.  Based on this, a search was done for interesting ones where
the copper ions sat on a frustrated lattice.  A partial list of these can be found in Table I.  The intent of this paper is to go through this
table and point out interesting materials that might be worth synthesizing and studying for their magnetic  properties.

\begin{table}
\begin{ruledtabular}
\begin{tabular}{llll}
Formula unit & SG 
&  Lattice & Ref \\
\colrule
Cu$_6$IO$_3$(OH)$_{10}$Cl & R3 & maple leaf & \cite{mills} \\
Cu$_6$TeO$_4$(OH)$_9$Cl	 & R3 & maple leaf & \cite{mills} \\
Pb$_3$Cu$_6$TeO$_6$(OH)$_7$Cl$_5$ & R$\bar{3}$ & maple leaf & \cite{kampf} \\
Cu$_2$ZnAsO$_4$(OH)$_3$ & P$\bar{3}$ & maple leaf & \cite{olmi} \\
Zn$_6$Cu$_3$(TeO$_6$)$_2$(OH)$_6$Y & P$\bar{3}$1m & kagome & \cite{burns} \\
MgCu$_2$TeO$_6$(H$_2$O)$_6$	 & P$\bar{3}$1m & honeycomb & \cite{margison} \\
Cu$_3$TeO$_6$(H$_2$O)$_2$ & P2$_1$/n & honeycomb/dimer & \cite{grice} \\
Cu$_3$TeO$_6$	 & Ia$\bar{3}$ & hexagons	 & \cite{falck1} \\
Pb$_3$(Cu$_5$Sb)$_{1/3}$(TeO$_3$)$_6$Cl & P4$_1$32 & hyperkagome & \cite{lam} \\
PbCuTe$_2$O$_6$ & P4$_1$32 & hyperkagome & \cite{kotes} \\
CuTeO$_4$	& P2$_1$/n & square lattice & \cite{falck2} \\
Sr$_2$CuTeO$_6	$ & I4/m & square lattice & \cite{iwanaga} \\
SrCuTe$_2$O$_7$ & Pbcm & orthorhombic & \cite{yeon} \\
Cu$_3$BiTe$_2$O$_8$Cl & Pcmn & kagome staircase & \cite{becker}
\end{tabular}
\end{ruledtabular}
\caption{Table of various copper tellurium oxides.  SG is the space group, Lattice is
the arrangement of copper ions, and Ref is the associated reference to the literature.}
\label{table1}
\end{table}

Bluebellite (Cu$_6$IO$_3$(OH)$_{10}$Cl) and mojaveite (Cu$_6$TeO$_4$(OH)$_9$Cl) are recently discovered minerals found in
the Mojave Desert \cite{mills}.  Though the first has I$^{5+}$ as opposed to Te$^{6+}$ for the second, their crystal structures are
similar, and also similar to another mineral discovered there, fuettererite (Pb$_3$Cu$_6$TeO$_6$(OH)$_7$Cl$_5$) \cite{kampf},
as well as the mineral sabelliite (Cu$_2$ZnAsO$_4$(OH)$_3$) \cite{olmi} discovered in Sardinia.  In these four examples, the copper ions
sit on a so-called maple leaf lattice (1/7-depleted triangular lattice).  This lattice (with a coordination number of z=5) is intermediate from
a frustration viewpoint between a triangular lattice (z=6) and a kagome (1/4-depleted triangular lattice with z=4), and is thought to be
(barely) on the ordered side \cite{schmal}.  There has been one copper mineral with such a lattice that has had its magnetic properties
investigated, spangolite (Cu$_6$Al(SO$_4$)(OH)$_{12}$Cl(H$_2$O)$_3$), whose susceptibility resembles that expected for a singlet
ground state \cite{fennell}, with a small upturn at low temperatures due to about a 7.5\% concentration of orphan spins (similar to what is
observed in herbertsmithite).  In all cases, though, the maple leaf lattice is distorted (Fig.~1).
In Table II, these distortions are tabulated, with sabelliite the least distorted, and bluebellite the most.
But for sabelliite, even though there is only one crystallographic Cu site (as compared to two for the others), significant site disorder exists in this material, with Zn on the Cu sites, and Sb on the As sites, making a synthetic variant a desirable goal.

\begin{table}
\begin{ruledtabular}
\begin{tabular}{lllll}
Formula unit & X 
&  S & L & L/S \\
\colrule
Cu$_6$IO$_3$(OH)$_{10}$Cl & I & 2.899 & 3.900 & 1.345 \\
Cu$_6$TeO$_4$(OH)$_9$Cl	 & Te & 2.999 & 3.572 & 1.191 \\
Pb$_3$Cu$_6$TeO$_6$(OH)$_7$Cl$_5$ & Te & 3.033 & 3.322 & 1.095 \\
Cu$_2$ZnAsO$_4$(OH)$_3$ & Zn & 3.028 & 3.166 & 1.046 \\
Cu$_6$Al(SO$_4$)(OH)$_{12}$Cl(H$_2$O)$_3$ & Al & 3.004 & 3.214 & 1.070
\end{tabular}
\end{ruledtabular}
\caption{Table of copper maple leaf lattices, 1/7-depleted triangular lattices with a sub formula unit of Cu$_6$X, with X sitting in
the middle of the hexagonal hole.  S is the smallest Cu-Cu near neighbor distance (in $\AA$), L the largest, with L/S their ratio.}
\label{table2}
\end{table}

\begin{figure}
\includegraphics[width=0.7\hsize]{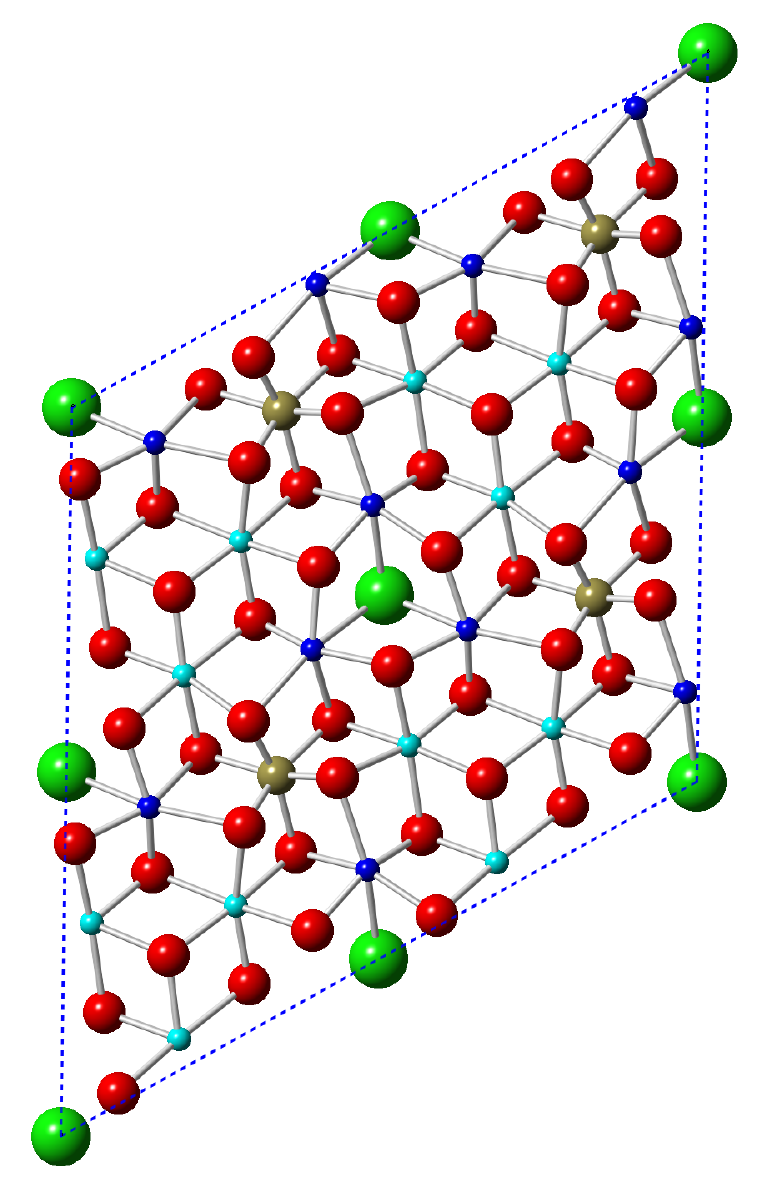}
\caption{Crystal structure of mojaveite \cite{mills}.  View is along $c$, with $z$ ranging from 0.2 to 0.5.  The two different crystallographic
Cu sites are shown as blue and cyan, with Cl as green, Te as gold and oxygen as red.}
\label{fig1}
\end{figure}

\begin{figure}
\includegraphics[width=0.7\hsize]{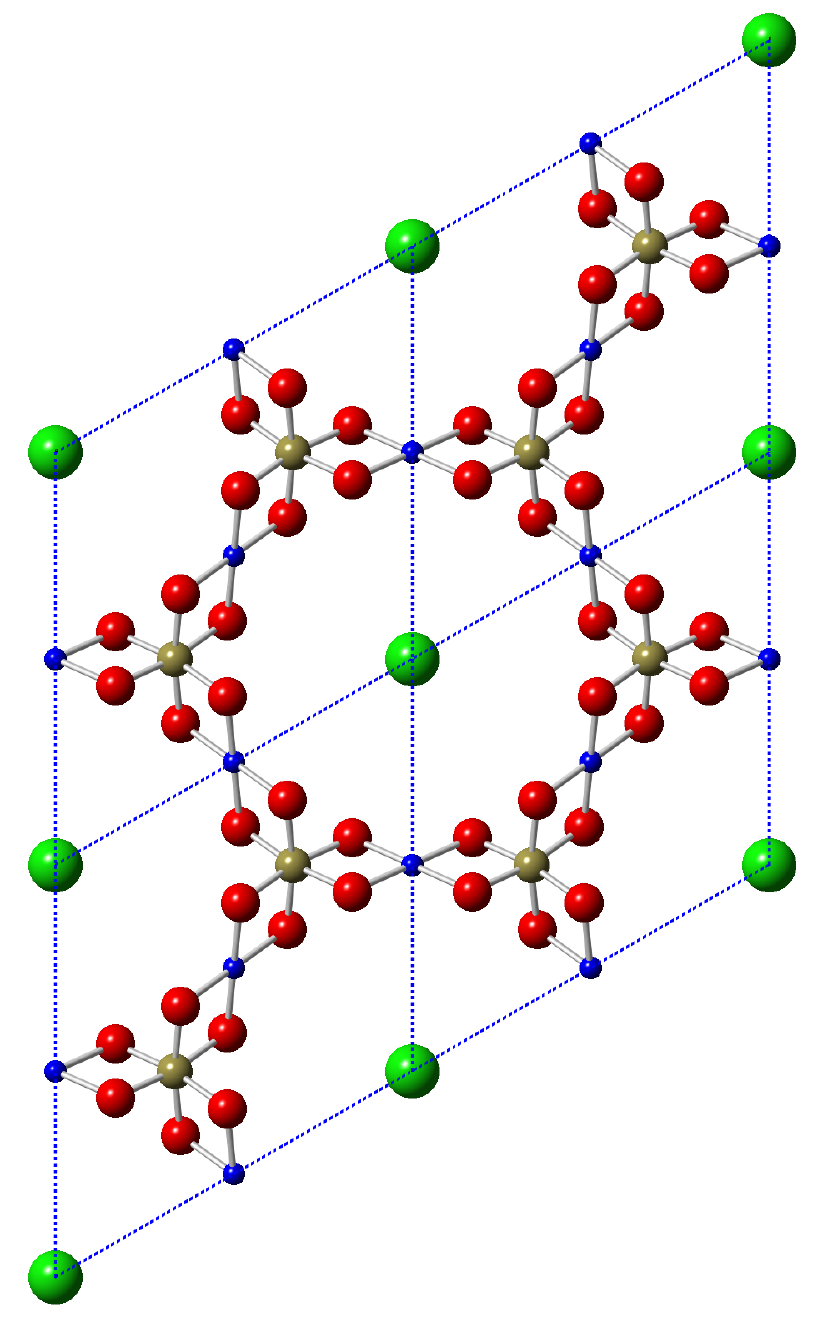}
\caption{Crystal structure of quetzalcoatlite \cite{burns}.  View is along $c$, with $z$ ranging from 0.25 to 0.75.  Cu is blue, Cl green, Te gold and oxygen red.}
\label{fig2}
\end{figure}

We next come to a more promising mineral, quetzalcoatlite (Zn$_6$Cu$_3$(TeO$_6$)$_2$(OH)$_6$Y,
with Y=Ag$_x$Pb$_y$Cl$_{x+2y}$ a neutral unit) \cite{burns}, found in the same Blue Bell claims as bluebellite.
Interestingly, this was the first mineral whose crystal structure was determined at the Advanced Photon Source at Argonne
(studied there because of the small size of the crystals).
It exhibits a perfect kagome net of copper ions (Fig.~2), with Cl ions sitting at the center of the hexagonal holes.
The stacking of the kagome planes is AA (as opposed to ABC stacking in herbertsmithite), thus similar to kapellasite, which
is a polymorph of herbertsmithite (both have the same space group).  Like herbertsmithite, these layers
are separated by Zn ions (Fig.~3), but here, the Zn ions have tetrahedral coordination (ZnO$_2$(OH)$_2$).  Because of this, the
disorder seen in herbertsmithite (where Cu can sit on the Zn sites) should be absent in this mineral.  Moreover, the Cu-O-Zn-O-Cu
pathway connecting successive kagome layers is quite tortuous, implying weak coupling between the layers.  But there is still
some disorder because of the variability of the Y unit (ideally one would have AgCl dimers along the $c$ axis).
Turning to the planar properties (Fig.~2), unlike herbertsmithite where one has Cu-O-Cu superexchange pathways, here Te intervenes,
leading to weaker couplings of the form Cu-O-Te-O-Cu or Cu-O-O-Cu (super-superexchange).  Because of this, the anticipated
Curie-Weiss temperature should be significantly smaller than herbertsmithite.  Still, as one of the few known materials where Cu ions fall
on a perfect kagome lattice, one would hope that the available crystals could be studied for their magnetic properties, and an
attempt at synthesis would be a desirable goal as well.

\begin{figure}
\includegraphics[width=1.0\hsize]{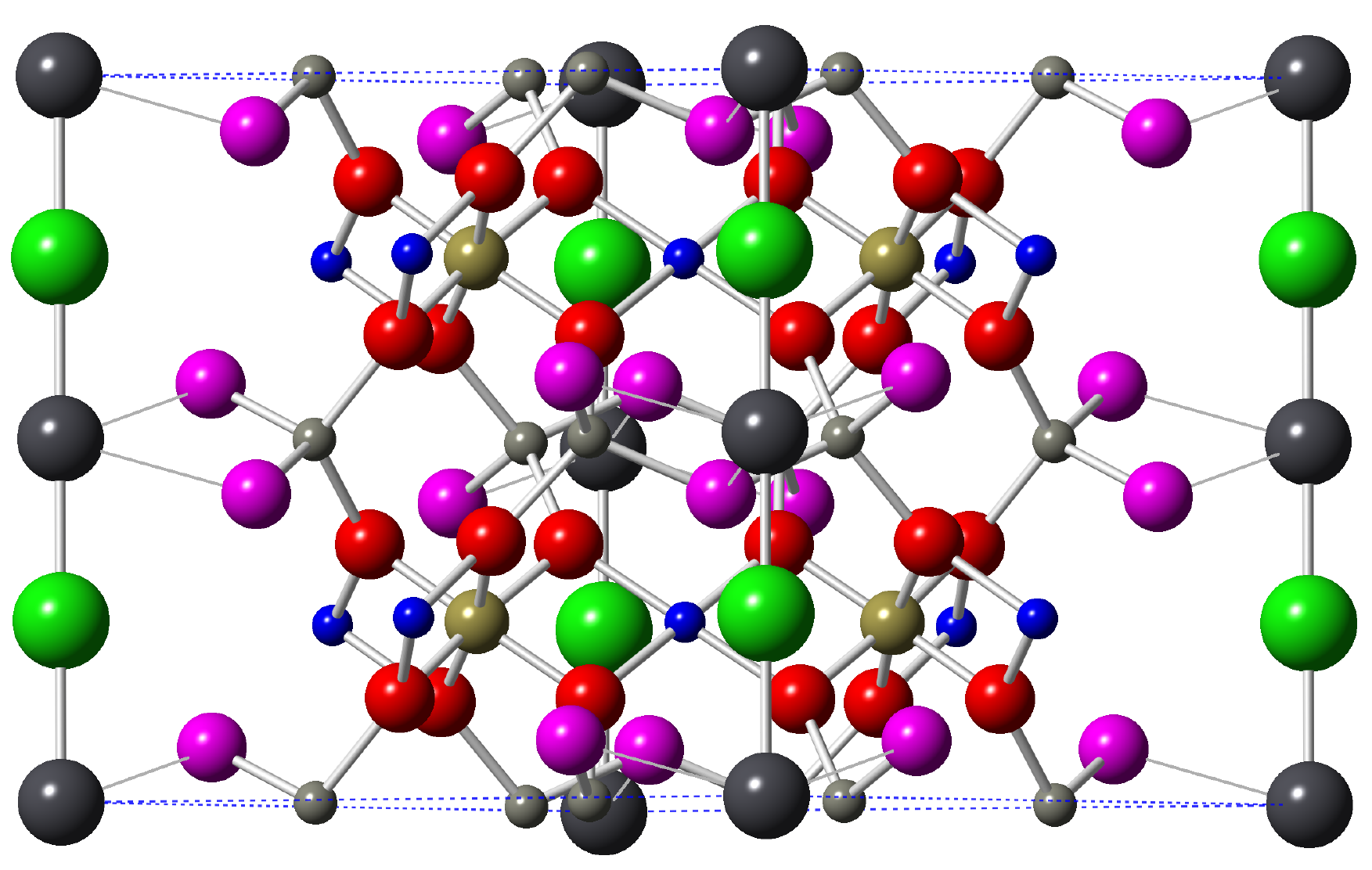}
\caption{Crystal structure of quetzalcoatlite \cite{burns}.  Shown is a side view (vertical axis along $c$).  Cu is blue, Cl green, Te gold, oxygen red,  Zn light gray, Pb dark gray, and OH groups pink (the positions of the hydrogens are not known).}
\label{fig3}
\end{figure}

\begin{figure}
\includegraphics[width=0.7\hsize]{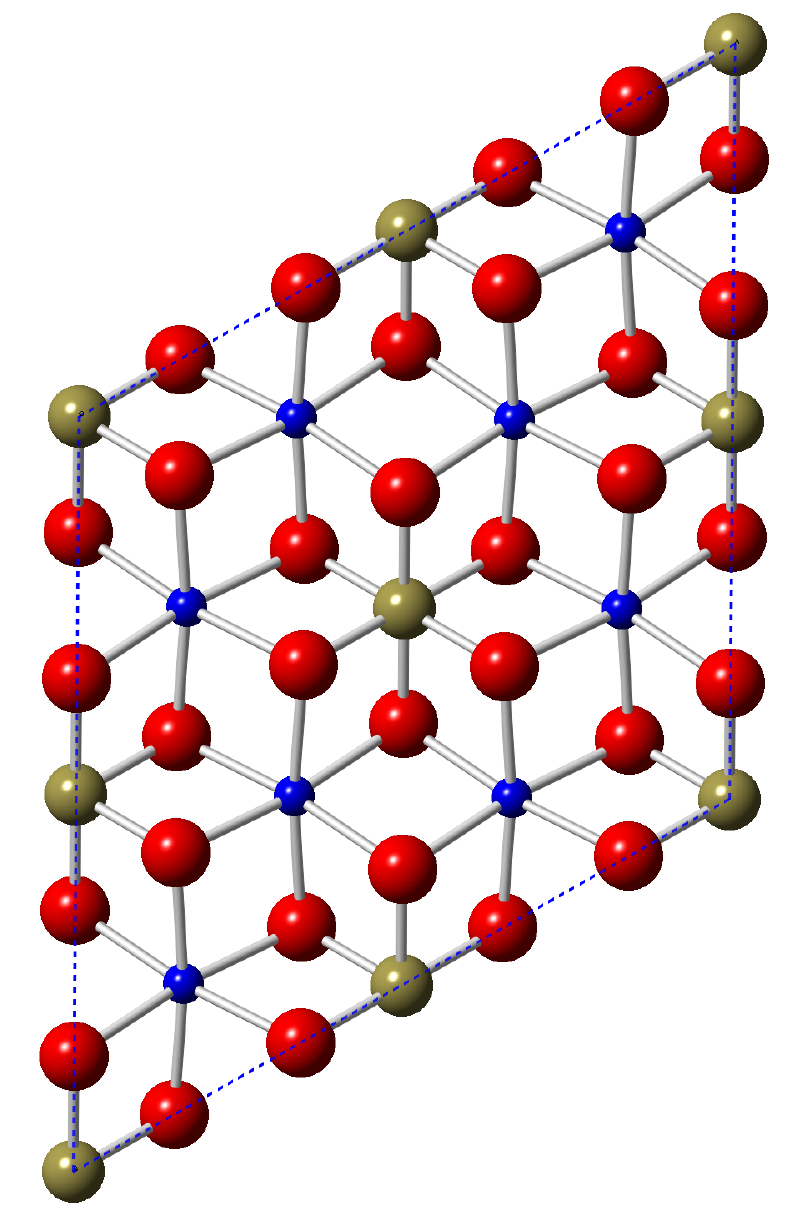}
\caption{Crystal structure of leisingite \cite{margison}.  View is along $c$, with $z$ ranging from 0.4 to 0.6.  Cu is blue, Te gold and oxygen red.}
\label{fig4}
\end{figure}

The next mineral in Table I is leisingite (MgCu$_2$TeO$_6$(H$_2$O)$_6$).  Here, the copper ions form a perfect honeycomb lattice (z=3),
with Te ions sitting in the hexagonal holes (Fig.~4).  The Cu ions are connected by a superexchange pathway, but the Cu-O-Cu bond
angle is 93.5$^\circ$, which is near the crossover from F to AFM behavior.  The layers are AA stacked, being conneced by Mg(H$_2$O)$_6$
octahedra with Mg ions sitting below the Te ions.  But again, disorder is present, with Fe sitting on the Mg sites, and some Mg sitting on the
Cu sites.  Again, a synthetic variant is highly desirable.

Jensenite (Cu$_3$TeO$_6$(H$_2$O)$_2$) is mentioned in passing.  Like leisingite, it is composed of layers of Te and Cu forming a
honeycomb lattice (but in this case, it is distorted) separated by other layers which contain isolated copper dimers.
Mcalpenite has the same formula unit as jensenite (except for the waters), but has a cubic space group instead.  Although known in mineral form \cite{carbone}, it has been synthesized as well by a variety of techniques  \cite{falck1,herak,calmi,he}.  It is composed of
a lattice of corner-sharing copper hexagons whose normals point in different directions  (a different lattice of hexagons has been seen in frustrated spinels
like ZnCr$_2$O$_4$ \cite{lee}).  This material, which has been studied by a number of groups, has been called a `spin web' compound \cite{calmi}
with a  N\'{e}el temperature of 61 K \cite{herak}.

If one takes a kagome lattice and then stretches it along the $c$ axis, one gets a hyperkagome structure, with corner sharing triangles arrayed
in a cubic space group.  The mineral choloalite (Pb$_3$(Cu$_5$Sb)$_{1/3}$(TeO$_3$)$_6$Cl) has this lattice \cite{lam}, with the same P4$_1$32 space group as the well known spin liquid iridate, Na$_4$Ir$_3$O$_8$ \cite{okamoto}.  A simpler synthetic material, PbCuTe$_2$O$_6$, has the same hyperkagome
lattice and space group \cite{kotes}.  No magnetic ordering has been seen by NMR and $\mu$SR down to 20 mK \cite{khuntia}, though thermodynamic
data indicate some type of transition occuring at 0.87 K \cite{kotes}, perhaps related to impurities \cite{khuntia}.  Because of the distoted nature of the lattice,
modeling this material is somewhat challenging, but exchange couplings have been proposed based on electronic structure  calculations \cite{kotes}.  A Sr variant is known that orders at 5.5 K \cite{ahmed}.

The first attempt to make Cu$_3$TeO$_6$ was by high temperature hydrothermal synthesis \cite{falck1}.  CuTeO$_4$ has been made under
similar conditions \cite{falck2}, though Cu$_3$TeO$_6$ is thermodynamically more stable.  The interest in CuTeO$_4$ is that it exhibits a
square lattice net for the copper ions (Fig.~5).  But unlike a typical cuprate, the Cu-O-Cu bond angles are either 122.5$^\circ$ of 126.1$^\circ$,
more simlar to herbertsmithite than other cuprates.  This buckling of the CuO$_2$ planes is due to an attempt to lattice match with a TeO$_2$
layer.  Interestingly, the copper ions are nearly octahedrally coordinated, with short, medium and long Cu-O bonds which at most are 14\%
different in length for Cu2 ions, and 18\% different for Cu1 ions.  Despite these differences from other cuprates, the electronic structure is remarkably similar, with a predicted magnetic ground state which is a quasi-2D N\'{e}el state \cite{botana}.  It has been proposed that replacing Te$^{6+}$
by Sb$^{5+}$ could hole dope this material, potentinally leading to a superconducting phase \cite{botana}.  Experimenally, only the structure is known,
and beyond the original synthesis paper \cite{falck2}, the only reports in the literature is finding it as a secondary phase.

\begin{figure}
\includegraphics[width=0.7\hsize]{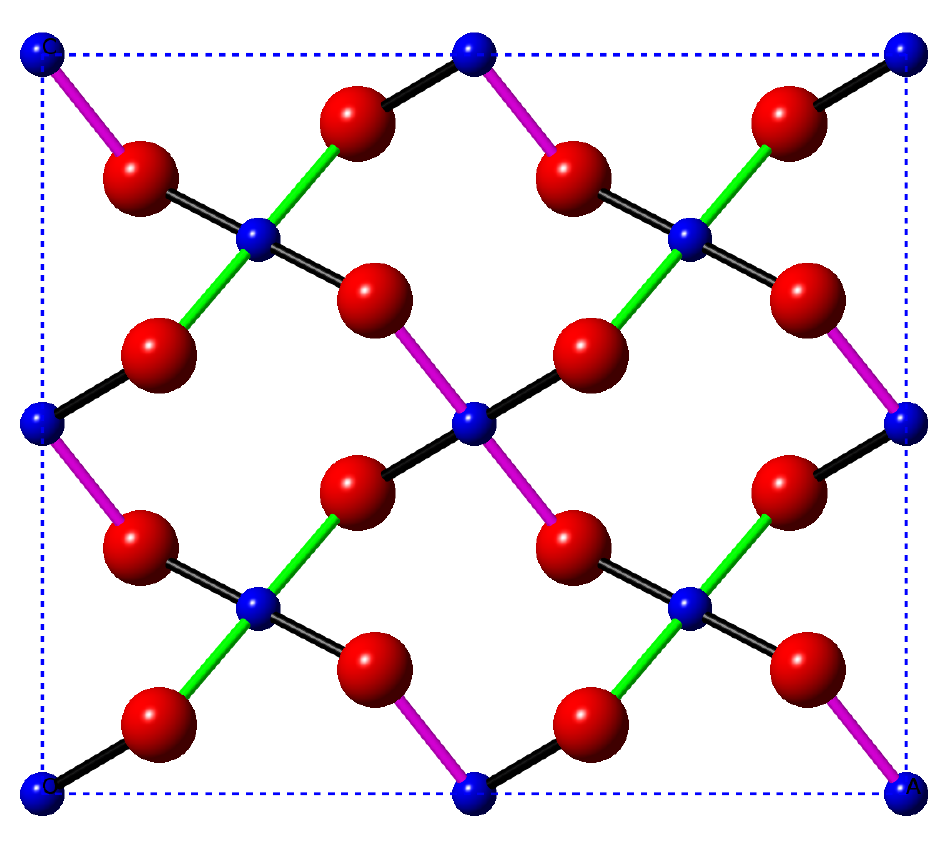}
\caption{Crystal structure of CuTeO$_4$ \cite{falck2}.  View is along the $b$ axis, with $y$ ranging from 0.4 to 0.6.  Cu is blue and oxygen red, with short
Cu-O bonds in black, medium bonds in green and long bonds in pink.}
\label{fig5}
\end{figure}

In the material Sr$_2$CuTeO$_6$ \cite{iwanaga}, the Te ions move down into the plane (Fig.~6).  So, though the copper ions again form a square net,
the exchange pathway is Cu-O-Te-O-Cu (or Cu-O-O-Cu) as in quetzalcoatlite.  This material is rather straightforward to synthesize, and a
number of variants are known (with Ba instead of Sr, or W instead of Te).  The Sr variant orders at $\sim$80 K \cite{iwanaga}, but the Ba
analgoue does not appear to order (for the W version, the Ba analogue orders at $\sim$30 K, but the Sr one does not).  These maetrials
have been extensively studied, as well as modeled by DFT calculations \cite{rosner}.  Crystals have been large enough to do both
elastic \cite{koga} and inelastic \cite{babkevich} neutron scattering.  One finds a simple N\'{e}el lattice, but with interactions significantly
reduced from the layered cuprates due to the super-superexchange nature of the magnetic coupling.

\begin{figure}
\includegraphics[width=0.7\hsize]{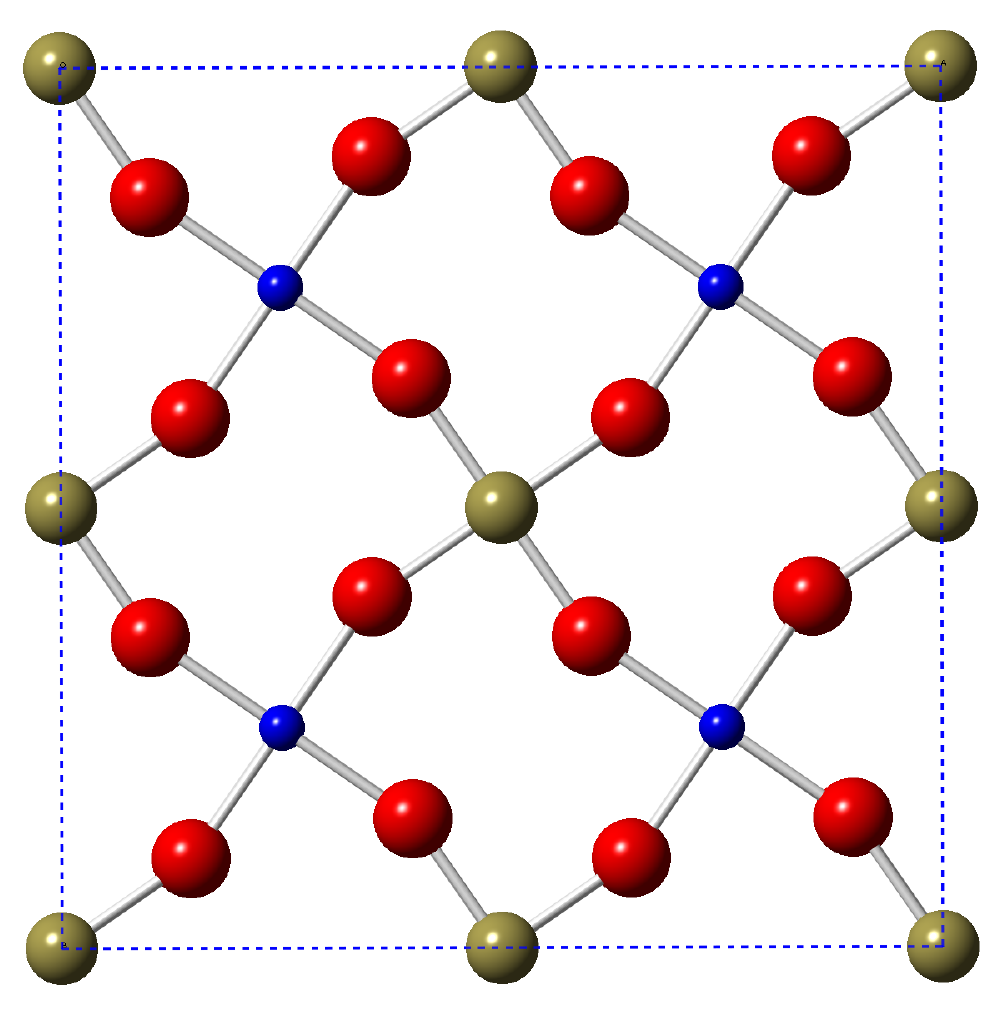}
\caption{Crystal structure of Sr$_2$CuTeO$_6$ \cite{iwanaga}.  View is along $c$, with $z$ ranging from 0.4 to 0.6.  Cu is blue, Te gold and oxygen red.}
\label{fig6}
\end{figure}

SrCuTe$_2$O$_7$ has two types of Te sites (one 4+, one 6+) \cite{yeon}.  The Cu ions sit on an orthorhombic lattice, either being
connected by a Cu-O-Sr-O-Cu pathway (long direction) or an orthogonal Cu-O-Te-O-Cu pathway (Fig.~7).  There are, though, four
different layers containing copper (Z=4).  No evidence for magnetic order has been found \cite{yeon}, but typically, an upturn is seen
in the susceptibility at low temperatures, indicating the presence of orphan spins.  Similar behavior is seen in Pb and Ba variants.

\begin{figure}
\includegraphics[width=1.0\hsize]{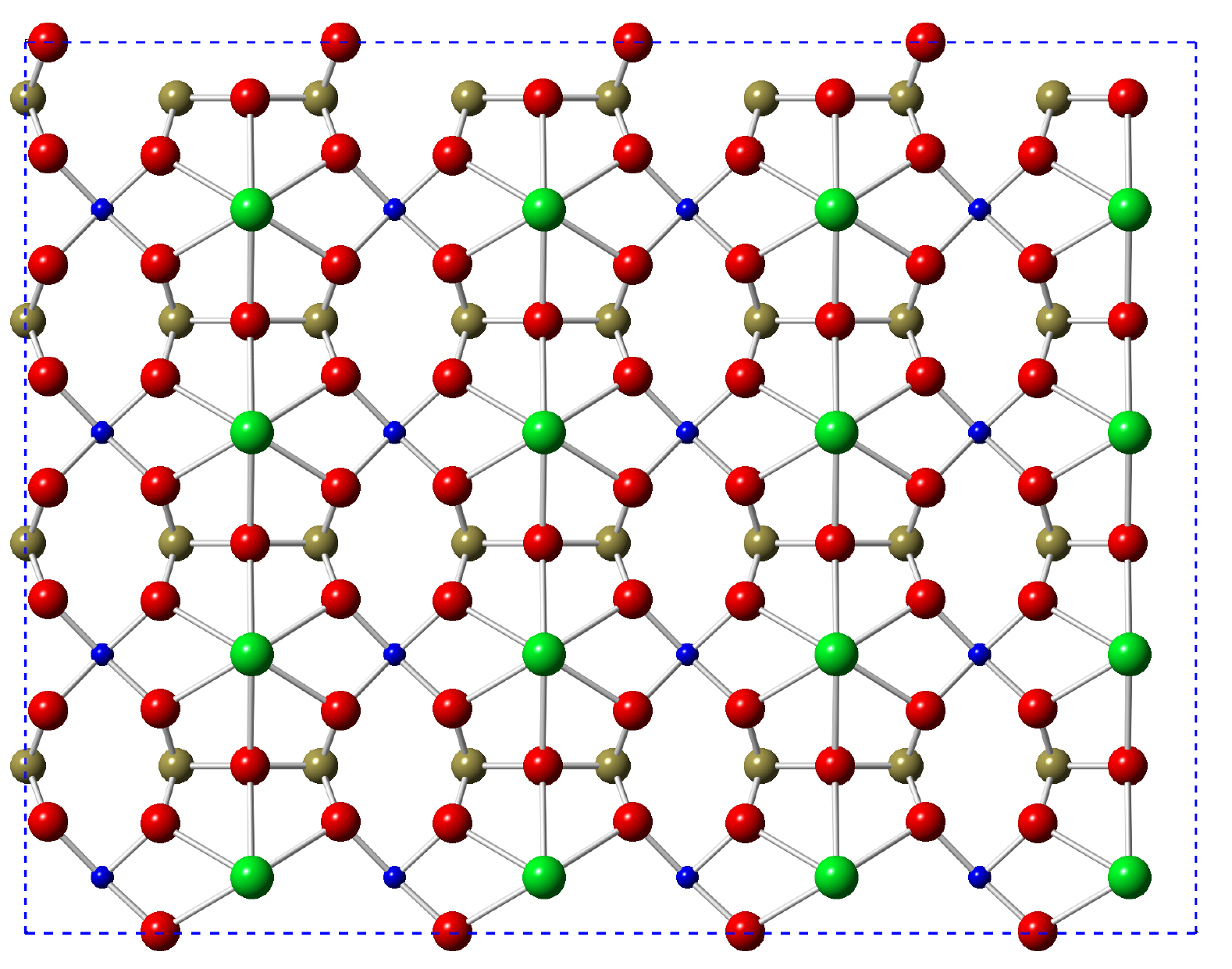}
\caption{Crystal structure of SrCuTe$_2$O$_7$ \cite{yeon}.  View is along the $b$ axis, with $y$ ranging from 0 to 0.2.  Cu is blue, Te gold and oxygen red, with Sr as green.}
\label{fig7}
\end{figure}

The last material we mention is Cu$_3$BiTe$_2$O$_8$Cl \cite{becker}, similar to the mineral francisite, Cu$_3$BiSe$_2$O$_6$Cl.
The lattice formed is a distorted version of a so-called kagome staricase, with the copper kagome layers strongly buckled.
Francisite itself has been synthesized and studied quite a bit \cite{millet,pregelj,rouso}, and orders at 27.4 K, but the order is
complicated due to the low symmetry of the lattice.

There are many other copper tellurium oxides that are not mentioned here, perhaps the best known being the perovskite CuTeO$_3$,
the Se variant of which is a ferromagnet \cite{subra}.  Interestingly, it has been recently modeled by Byung Il Min's group \cite{min}, he being a former student of Art's that I had the pleasure of working with when I was a postdoc of Art's.
Many materials not mentioned here exhibit chains instead, or more complex
lattices.  Variants are also known where Sb replaces Te.  Certainly, Mother Nature has been kind to us in providing a wonderful
playground of unexplored materials.  It is up to us to do the exploring.

I know that if Art were still with us, he and his group would likely take on the challenge of trying
to better understand this fascinating class of materials.  Ironically, the motivation behind the study of
spin liquids is the original work of Phil Anderson as outlined at the beginning of this article, and the two of them were well known
for not seeing eye to eye (Art being a student of Slater's, and Phil of van Vleck's).  In that context, I cannot resist showing a cartoon I presented to Art on his 80$^{th}$ birthday
(Fig.~8) motivated by a quote from Martin Peter that succinctly illustrates the complex relation between Art and Phil \cite{peter}.
I should end by saying that the physicist I came to be was shaped by the melding of the influences that these two
individuals had on me, both of which I am eternally grateful to.

\begin{figure}
\includegraphics[width=0.9\hsize]{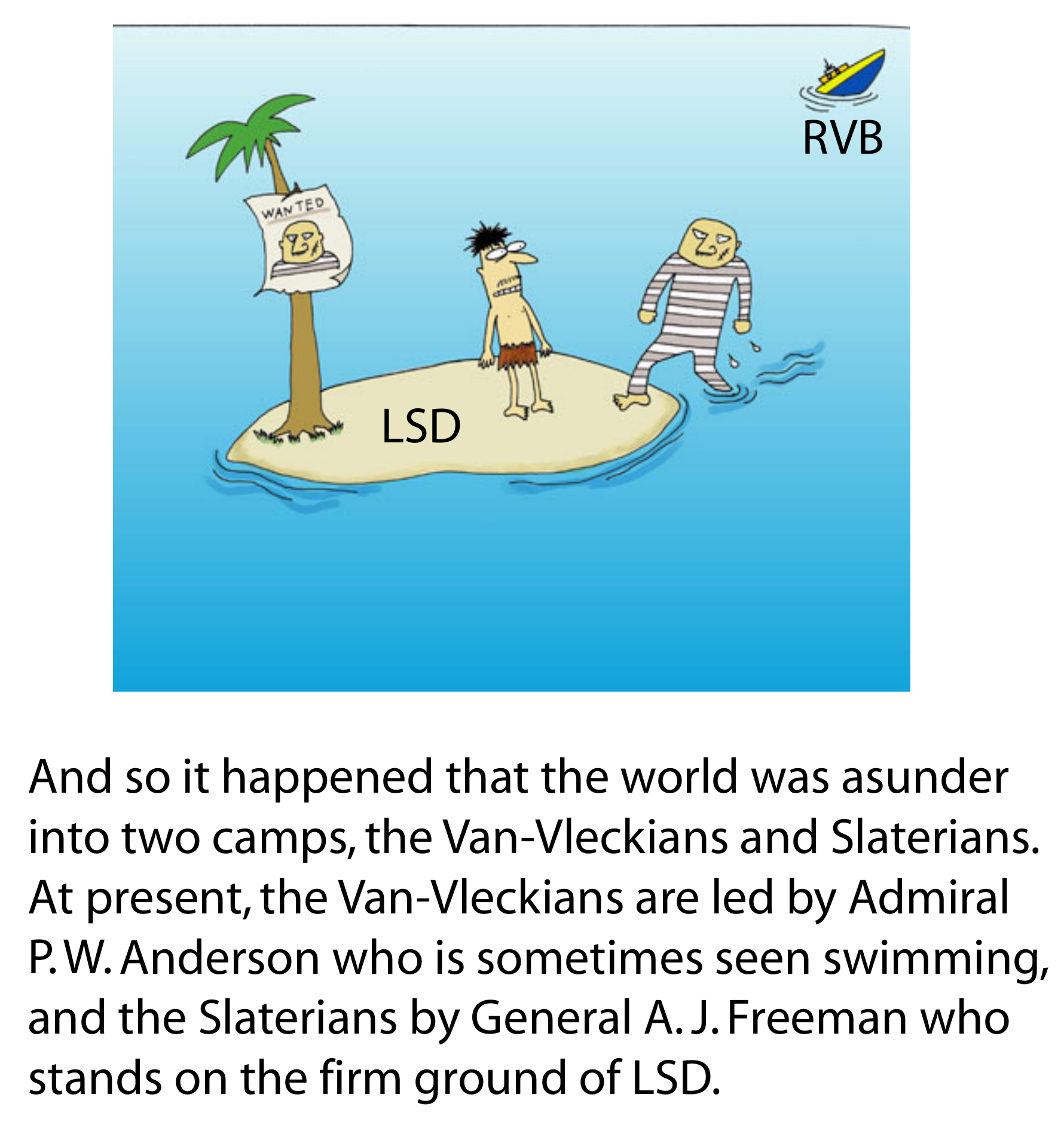}
\caption{Cartoon presented by Mike Norman to Art Freeman on his 80$^{th}$ birthday.  The quote is from Martin Peter \cite{peter}.}
\label{fig8}
\end{figure}

Work supported by the Materials Sciences and Engineering Division, Basic Energy Sciences, Office of Science, US DOE.
The author would like to thank Antia Botana, who was instrumental in the work on CuTeO$_4$, and as an aside, is a `grandchild' of Art's
(having been supervised as a postdoc by two people who were supervised as postdocs by Art, myself and Warren Pickett).
And my thanks to Ruqian Wu, Bruce Harmon and Sam Bader for this opportunity to honor Art's memory.

\end{document}